\def\unib{UNi$\rm_4B$ }
\def\ube13{UBe$\rm_{13}$}

\documentstyle[aps,prl,twocolumn,epsf]{revtex}
\begin{document} 
\draft

\def\dfrac#1#2{{\displaystyle{#1\over#2}}}
\twocolumn[\hsize\textwidth\columnwidth\hsize\csname @twocolumnfalse\endcsname
Submitted to Phys. Rev. Lett. \hfill{LA-UR-99-143}

\title{Second Low Temperature Phase Transition in Frustrated UNi$\rm_4$B}



\author{R. Movshovich and  M. Jaime} 
\address{Los Alamos National Laboratory, Los Alamos, New Mexico 87545}

\author{S. Mentink,$^*$ A. A. Menovsky, and J. A. Mydosh}                                                                                            
\address{Kamerling Onnes Laboratory, Leiden University, P.O. Box 9504, 2300 RA Leiden, The Netherlands}

\date{\today}

\maketitle

\begin{abstract} 

Hexagonal UNi$\rm_4$B is magnetically frustrated, yet it orders antiferromagnetically at T$\rm_N$ = 20 K. However, one third of the U-spins remain paramagnetic below this temperature. In order to track these spins to lower temperature, we measured the specific heat $C$ of \unib between 100 mK and 2 K, and in applied fields up to 9 T. For zero field there is a sharp kink in $C$ at $T^\ast\approx$ 330 mK, which we interpret as an indication of a second phase transition involving paramagnetic U. The rise in $\gamma = C/T$ between 7 K and 330 mK and the absence of a large entropy liberated at $T^\ast$ may be due to a combination of Kondo screening effects and frustration that strongly modifies the low $T$ transition. 

\end{abstract}


]
\narrowtext

\unib has been an object of intense experimental and theoretical study in the last several years.\cite{mentink:prl_94,mentink:prb_95,lacroix:prl_96} The main reason for such interest is a highly unconventional antiferromagnetically ordered state that this compounds attains below the phase transition temperature of $T_N = 20$ K.  Only 2/3 of the U atoms order, with the rest remaining paramagnetic below $T_N$.\cite{mentink:prl_94} The origin of such behavior must be sought in the frustrating nature of the triangular crystallographic lattice  in which \unib forms. 

\begin{figure}
\epsfxsize=2.6in
\centerline{\epsfbox{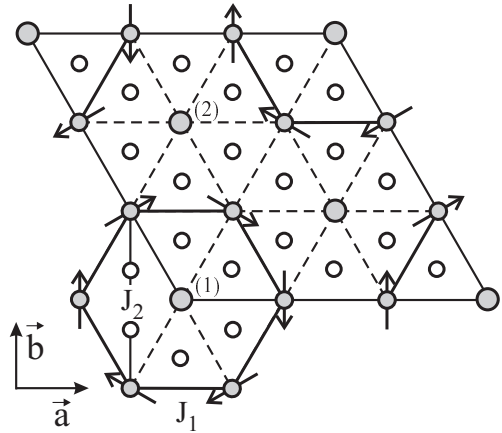}}
\caption{Magnetic structure of \unib in $a - b$ plane, from Ref. [1]. U atoms at sites (1) and (2) remain paramagnetic below $T_N$. Open circles: Ni or B atoms.} 
\label{structure}
\end{figure}

The crystal structure of \unib corresponds to the hexagonal CeCo$\rm_4$B-type.~\cite{mentink:physica_93} The U- and Ni- (or B-) containing triangular planes are shown in Fig.~\ref{structure}. Within these planes both nearest (nn) and next nearest neighbors (nnn) interactions are antiferromagnetic, with $a-b$ an easy magnetization plane. Below $T_N$ this highly frustrated system partially orders, with magnetic unit cell containing nine U atoms. Six of them form an in-plane vortex-like pattern, with neighboring U spins rotated by $60^\circ$. The remaining three U atoms remain paramagnetic and occupy two distinct positions: one is in the center of the vortex; two other are between the vortices and are surrounded by three pairs of antiparallel ordered U spins. The U atoms are coupled ferromagnetically along the $c$-axis, creating in 3D an ordered array of ferromagnetic and paramagnetic chains. 

A number of transport and thermodynamic properties were measured to investigate the ordered phase of UNi$\rm_4$B, both in zero~\cite{mentink:prl_94} and applied magnetic field.~\cite{mentink:prb_95,mentink:physicaB_97} Resistivity in the $a-b$ plane continues to rise below $T_N$, peaks at 5 K, and then drops rather sharply. The total range of variation in resistance over this temperature range is small, about 4\%. The specific heat divided by temperature $C / T = \gamma $ initially drops below $T_N = 20$ K, and starts rising again below 7 K. $\gamma$  continues to rise down to the lowest previously measured temperature of $T = 0.35$ K to $\approx 0.5$ $\rm J/mol K^2$. Application of magnetic field up to 16 T suppressed $\gamma$ by about a factor of 3. These results were taken as an indication that Kondo effect plays an important role in determining the low temperature properties of UNi$\rm_4$B.~\cite{lacroix:prl_96,mentink:physicaB_97}
 
Several theoretical attempts were made to reproduce the unique partially ordered state below $T_N$ and interpret the low temperature specific heat. Initially,~\cite{mentink:prl_94} ferromagnetic fluctuations in the paramagnetic 1D chains were suggested to explain the low temperature upturn in $\gamma$. The specific heat calculated for a 1D Heisenberg ferromagnetic chain~\cite{bonner:64} with $S = 1/2$ and $J_c = 35 $ K gave a rather good representation of the measured low temperature tail in specific heat. An alternative view point was taken by Lacroix {\it et al.} (Ref.~\onlinecite{lacroix:prl_96}), where a model was developed to treat both geometric frustration and a possible Kondo interaction between the paramagnetic U spins and conduction electrons. The starting point of this model postulates that the 1D U chains in the $c$-axis

\begin{figure}
\epsfxsize=3.in
\centerline{\epsfbox{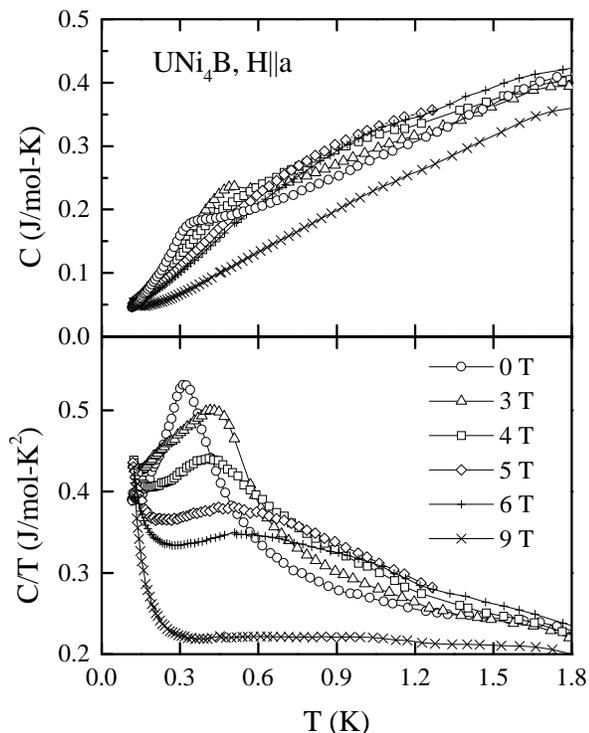}}
\caption{(a) Specific heat of \unib in magnetic field with $\vec{H} \parallel \vec{a}$. ($\circ$) $H = 0$ T; ($\triangle$) $H = 3$ T; ($\Box$) $H =  4$ T; ($\Diamond$) $H =  5$ T; ($+$) $H = 6$ T; ($\times$) $H = 9$ T. (b) Specific heat divided by temperature ($\gamma$) for the data from (a).} 
\label{hc:h_par_a}
\end{figure}

\noindent  are close to magnetic-nonmagnetic instability between the ferromagnetic alignment of the U-spins and a 1D lattice of Kondo-screened zero spin U atoms. Within this model several ground states are possible depending on the strength of the nn and nnn exchange interactions ($J_1$ and $J_2$, respectively) as well as the energy $\Delta$ (taken positive in the model) to create a magnetic chain and overcome Kondo screening. For sufficiently small values of $J_1$ and $J_2$  Kondo effect dominates and results in a non-magnetic (NM) phase, with all U spins Kondo-compensated. In the intermediate range of $J_1$ and $J_2$, with slight experimentally observed distortion taken into account,~\cite{drost:thesis,mentink:physicaB_96} the stable structure is the observed mixed phase described above. To stabilize a different ground state, additional interaction, perhaps due to crystallographic distortion that makes $J_1$ and $J_2$ anisotropic, is required within this model.~\cite{nunez-regueiro:private}

Another approach treats U's  as classical Heisenberg spins in $a - b$ plane.~\cite{tejima:97} Again, nn and nnn interactions are taken into account, as well as interhexagon exchange coupling. For the appropriate choices of parameters, quantum fluctuations can destabilize the standard 120$^\circ$ (3 sublattice) Ne\'{e}l order, and minimization of the total energy gives the experimentally observed ground state. The calculated $\gamma$ has a broad maximum at 2 K, and smoothly decreasing to zero as $ T \rightarrow 0$, due to the dominant contribution of spin waves. Therefore, this model seams unable to reproduce experimentally observed specific heat. One must, however, consider the possibility that the low temperature anomaly, due to disordered 1/3 of U atoms, is superimposed upon such spin-wave behavior.

It should be possible to distinguish between these scenarios by performing specific heat measurement at lower temperatures and compare the data with detailed predictions of the 1D ferromagnetic chain and the Kondo models. This was the original motivation behind the measurements that are the subject of this Letter. Our zero-field heat capacity data collected down to 100 mK showed a sharp kink at a temperature of $ \approx 330$ mK, not predicted by any of the theories discussed above. Further measurements in magnetic field up to 9 Tesla parallel to $a$- or $b$-axis revealed a very unusual evolution of this feature with field. We propose that this anomaly is an indication of the phase transition involving the U spins that remain paramagnetic below $T_N = 20 $ K. We suggest that the results described below are qualitatively consistent with the model of Kondo-screening of the U spins in the non-magnetic chains,~\cite{lacroix:prl_96} which severely affect their transition at $T^\ast$.

The single crystal of \unib used in this experiment (with a mass of $\approx 173$ mg) was grown with Czochralski technique. Similarly produced samples were evaluated with microprobe analysis and neutron diffraction, and were found to be of high quality~\cite{mentink:prl_94} (no second phase and without disorder). Specific heat data were collected with a quasiadiabatic technique,~\cite{quas_adb} where ruthenium oxide thick film resistors~\cite{sota} were used for thermometry. These resistors were previously calibrated as a function of temperature in a magnetic field against a thermometer placed in a field-free region of the apparatus. 

\begin{figure}
\epsfxsize=3.in
\centerline{\epsfbox{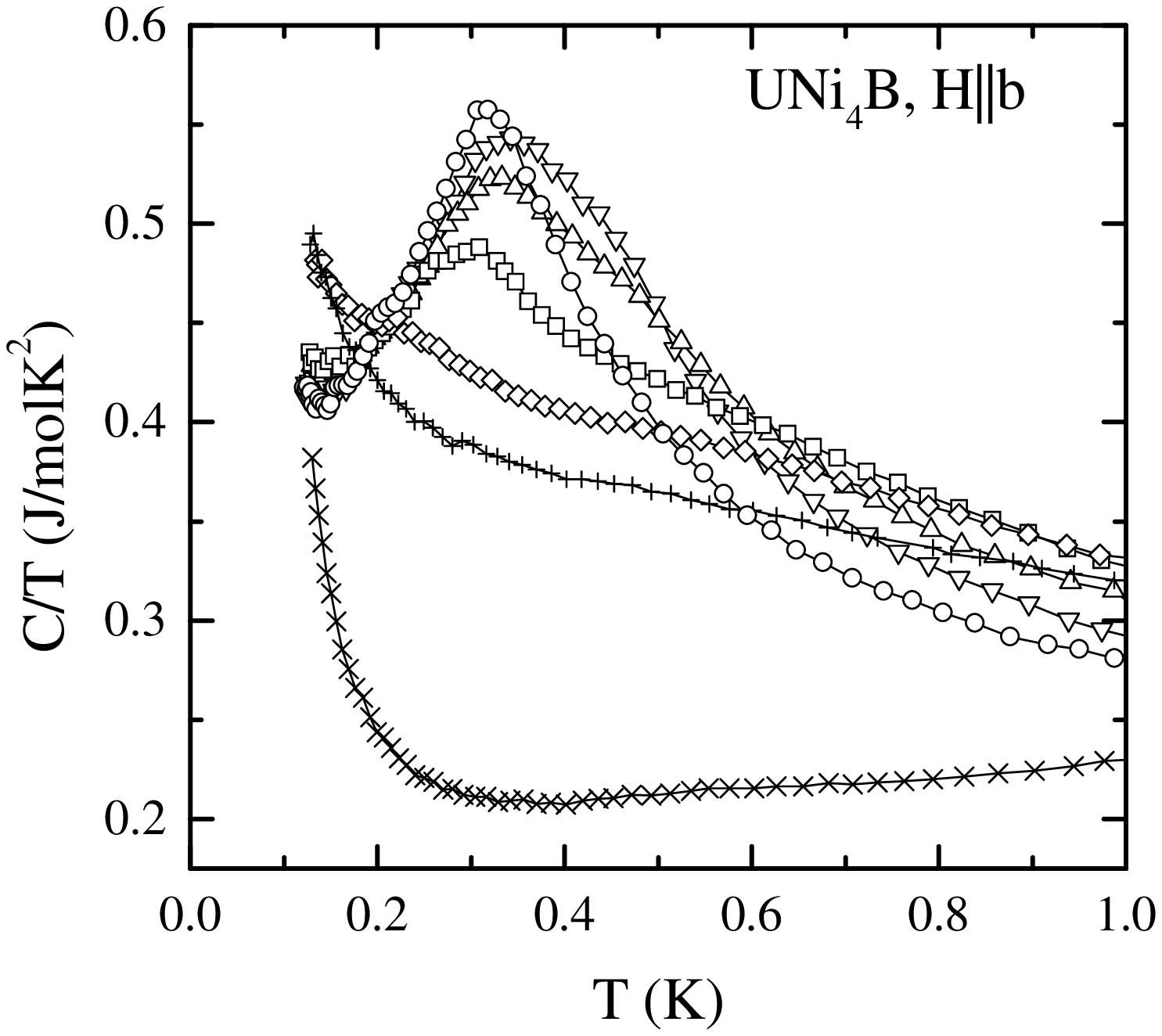}}
\caption{Specific heat divided by temperature of \unib in magnetic field with $\vec{H} \parallel \vec{b}$. ($\circ$) $H = 0$ T; ($\bigtriangledown$) $H = 2$ T; ($\triangle$) $H = 3$ T; ($\Box$) $H =  4$ T; ($\Diamond$) $H =  5$ T; ($+$) $H = 6$ T; ($\times$) $H = 9$ T.}
\label{hc:h_par_b}
\end{figure}

Fig.~\ref{hc:h_par_a} shows the specific heat data collected with magnetic field parallel to the $a$-axis (along the line connecting nearest in-plane U neighbors), where we plot both specific heat (a) and $\gamma = C/T$ (b) for magnetic field up to 9 Tesla. Not all available field data are shown in the figure for the sake of clarity. The anomaly in zero field appears as a clear kink in the specific heat at a temperature of $\approx 330$ mK. Just below this temperature the rise in $\gamma$ is interrupted, as shown in Fig.~\ref{hc:h_par_a}(b). Application of magnetic field initially moves the anomaly to higher temperature, with the temperature $T^\ast$ of the peak in $\gamma$ reaching  maximum at about 3 T. For still larger fields the anomaly
first broadens, and then the broad hump in $\gamma$ shifts to higher temperatures for fields above 4 T. The field of 9 T completely destroys the  anomaly, yet a pronounced low temperature tail emerges, which is present at all measured fields. The apparent insensitivity of the lowest temperature specific heat to magnetic field of up to 9 T indicates that the low temperature tail is likely due to the nuclear Schottky anomaly in large internal fields produced by the ordered U spins. Fig.~\ref{hc:h_par_b} shows specific heat data collected with the field parallel to the $b$-axis (along the line connecting the U next nearest neighbors), 

\noindent where we plot only $\gamma$ vs. temperature. The peak initially moves slightly to higher temperature for fields up to 2 T, before turning around, and is shifted to zero by the field of 6 T. As in the case of $\vec{H} \parallel \vec{a}$, the field of 9 T completely suppresses the anomaly, and displays a low temperature Schottky-like tail. 

\begin{figure}
\epsfxsize=3.in
\centerline{\epsfbox{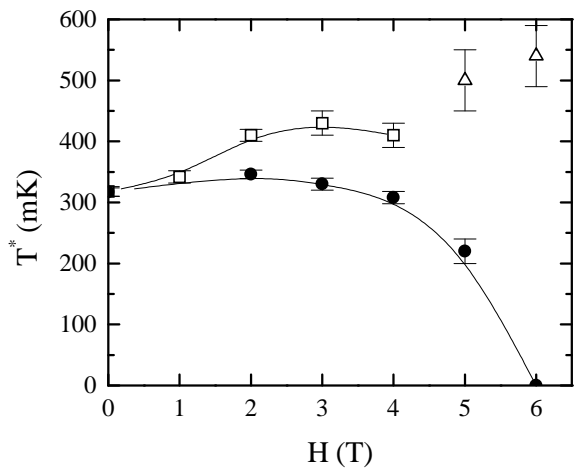}}
\caption{Phase transition temperature $T^\ast$ vs. field. ($\Box$, $\triangle$) $\vec{H} \parallel \vec{a}$; ($\bullet$) $\vec{H} \parallel \vec{b}$. Solid lines are guides to the eye.}
\label{t_star}
\end{figure}

To compare the data for different field orientations, we plotted $T^\ast$ as a function of magnetic field along both $a$- and $b$-axis in Fig.~\ref{t_star}. For the field along the $a$-axis the dependence is not monotonic, with a break at 4 T. Low temperature magnetic susceptibility measurements performed at 200 mK as a function of magnetic field in the same orientation ($\vec{H} \parallel \vec{a}$) show a change in slope (a kink) at a field of 4 T,~\cite{meisel:private} perhaps indicating spin reorientation, or crossover as observed in magnetoresistance.~\cite{mentink:thesis} It is likely that the break in the behavior of $T^\ast$ vs. field at 4 T for $\vec{H} \parallel \vec{a}$ is related to the same phenomenon. For both $\vec{H} \parallel \vec{a}$ and $\vec{H} \parallel \vec{b}$, $T^\ast$ initially rises with field, though this feature is much more pronounced for $\vec{H} \parallel \vec{a}$. For $\vec{H} \parallel \vec{b}$ orientation $T^\ast$ is suppressed smoothly to zero by the field of 6 T, indicating the absence of the spin reorientation phenomena for this range of field and its orientation. Spin reorientation (metamagnetic) transitions in \unib have been observed in 
the past via both magnetization and resistivity measurements,~\cite{mentink:prb_95} with the zero-field structure more resilient to the field applied in $b$- than in $a$-direction. One of the very surprising features of the ordered phase of \unib below $T_N = 20$ K was the absence of subsequent ordering of the U spins in paramagnetic chains. These chains are coupled by the $J_2$ exchange interaction which appears to be dominant in the $a-b$ plane. This strong interaction would be expected to drive the ordering of the paramagnetic chains as the temperature is lowered further below $T_N$.  There
  
\noindent are other examples of magnetic systems that display cascade of ordering transitions,both insulating and itinerant.~\cite{ramirez:arms_94,lacroix:physicaB_97,mydosh:zpb_97} The insulating Ising triangular system CsCoBr$\rm_3$ undergoes the first phase transition at 28 K, where, just as in the case of UNi$\rm_4$B, only 2/3 of the spins participate, with remaining 1/3 of the spins ordering antiferromagnetically at 12 K, a temperature three times lower.~\cite{yelon:prb_75,farkas:jap_91} In the case of \unib we can say now that the ordering does take place as well. However, the difference between the temperatures of the two observed phase transition  in \unib is much greater, a factor of 60. Yet, we expect the ferromagnetic coupling $J_c$ along the chains and the antiferromagnetic exchange interaction $J_2$ in $a$-$b$ planes to drive both high and low temperature phase transitions. We believe that the origin of the large difference between the ratios of the phase transition temperatures in the two systems lies in the fact that CsCoBr$\rm _3$ is an insulator and \unib is a metal. Kondo screening of the paramagnetic U spins by conduction electrons in \unib play crucial role in suppressing the second antiferromagnetic ordering temperature $T^\ast$. 

Within this scenario we can understand several features of the specific heat data. First is the size of the anomaly in specific heat associated with the low temperature phase transition. We do not see a step that would indicate a second order mean-field magnetic phase transition. Instead, it is manifested by a kink in the specific heat. The amount of entropy released at the low phase transition temperature $T^\ast$ is $0.1$ $\rm J / mol K$, forty times lower than $0.72 R \ln 2 = 4.15$ $\rm J / mol K$ of magnetic entropy recovered at 25 K.~\cite{mentink:physica_94,mentink:thesis} At 2 K the entropy grows to $0.57$ $\rm J / mol K$, close to 30\% of the $1/3 R \ln 2$ of the total entropy one can expect for the U spins in the paramagnetic chains (if the ground state is a doublet). There are two mechanisms at work that result in small amount of entropy liberated by the lower transition. (i)Frustration affects both the low and high temperature ordering transitions. The strongest interaction that couples U spins is ferromagnetic exchange $J_c$ along the $c$-axis. The frustration in the $a-b$ plane prevents the ferromagnetic ordering from taking place at the mean-field temperature corresponding to $J_c$. As a result, $T_N$ is depressed and substantial entropy is released above $T_N$ in via ferromagnetic fluctuations along the $c$-axis. Below $T_N$ paramagnetic sites are not equivalent, some being in the center of the ordered vortices, position (1), and some being between the vortices, position (2). However, magnetic field from the ordered U spins cancels at both sites, and the remaining U atoms can be viewed as triangular lattice with nn exchange interaction $J_2$. Therefore, as the temperature is lowered further, U spins in paramagnetic chains experience the frustration inherent to the triangular lattice. (ii)Kondo screening is present with characteristic temperature $T_K \approx 9$ K.~\cite{mentink:thesis,mentink:physicaB_97} Such screening with this $T_K$ alone would be expected to effectively reduce the U spins at $T^\ast $, absorbing most of the spin entropy. Frustration (or ferromagnetic fluctuations along the $c$-axis) in addition absorbs more entropy above $T^\ast$. As a result, most of the entropy associated with paramagnetic U spins  is liberated well $T^\ast$, resulting in a small specific heat feature at the transition. 

Secondly, the very unusual evolution of $T^\ast$ with magnetic field, displayed in Fig.~\ref{t_star}, may also have its explanation in the Kondo screening of the paramagnetic U spins. Magnetic field is expected to break Kondo singlets. The increased spins on paramagnetic sites would tend to order at a higher temperature. The reversal of this trend at higher magnetic field (especially pronounced for $\vec{H}\parallel \vec{b}$ orientation) is most likely due to the usual tendency of the magnetic field to suppress the antiferromagnetic order.

Other scenarios may be invoked to explain the low temperature anomaly in specific heat at 330 mK.  One possibility is spin reorientation transition involving the spins that are ordered below $T_N$.  This scenario may be at odds with calculations of Ref.~\onlinecite{lacroix:prl_96} which gives only one configuration with 2/3 of the U spin participating in antiferromagnetic ordering as stable ground state. Perhaps under some conditions other phases can be stabilized, involving spin reorientation. Another possibility is that frustration drives a spin-glass-like freezing of paramagnetic U spins. This particular scenario agrees with the observed initial effect of the magnetic field (up to 3 T, see Fig.~\ref{t_star}), since applied field tends to increase the temperature $T_f$ of the spin-glass freezing. Finally, there is a possibility that the low temperature anomaly represents a cross-over into a new quantum state of spins, resembling a quantum liquid, since AC-susceptibility~\cite{meisel:private} and preliminary $\mu$SR experiments~\cite{nieuwenhuys:private} show featureless magnetic response and no rise of onset of static, non-random internal magnetic field. These results would be also consistent with a small moment on almost perfectly screened and ordered U spins.  Sensitive neutron scattering and additional $\mu$SR experiments could be very useful in testing our suggestion for the origin of the observed specific heat anomaly.

In conclusion, we have discovered a second low temperature phase transition in the magnetically frustrated triangular \unib. The low temperature $T^\ast = 330$ mK of the second transition, very large ratio $T_N / T^\ast = 60$, and a non-monotonic evolution of the specific heat anomaly with magnetic field can be qualitatively explained with a combination the effects of Kondo-screening~\cite{lacroix:prl_96} and geometric frustration. 

We acknowledge helpful conversations with M. Meisel and G. J. Nieuwenhuys, and thank them for making available to us their unpublished data.

Work at Los Alamos was performed under the auspices  of the Department of Energy. Part of this research was supported by the Dutch Foundation FOM.


\end{document}